\def \be {\begin{equation}}
\def \ee {\end{equation}}
\def \ts {\textstyle}
\def \ni {\noindent}

\def \bea {\begin{eqnarray}}
\def \eea {\end{eqnarray}}
\documentstyle[12pt]{article}

\begin{document}
\begin{flushright}
{\sf Portsmouth University \\
Relativity and Cosmology Group \\
{\em Preprint} RCG 95/06}
\end{flushright}
\begin{center}
\[ \]
{\Large \bf Almost--homogeneity of the Universe in Higher--order
Gravity}
\[ \]
David R Taylor\footnote{Dept. Computational and Applied Mathematics,
Witwatersrand University, 2050, South Africa}${}^,$\footnote
{School of Mathematics, Portsmouth University,
PO1 2EG, England}${}^,$\footnote{Centre for
Nonlinear Studies, Witwatersrand University, 2050, South Africa}
and Roy Maartens${}^{2,3}$
\[ \]
{\bf Abstract}\\
\end{center}

\noindent In the $R+\alpha R^2$ gravity theory, we show that
if freely propagating massless
particles have an almost isotropic distribution, then the spacetime
is almost
Friedmann--Robertson--Walker (FRW). This extends the
result proved recently in general relativity
($\alpha=0$), which is applicable
to the microwave background after photon decoupling.
The higher--order result is in principle applicable to a massless
species that decouples in the early universe, such as a
relic graviton background.
Any future observations that
show small anisotropies in such a background would imply
that the geometry of the early universe were almost FRW.
\[ \]
\section{Introduction}

Recently it has been shown [1] that the observed almost--isotropy of
the
cosmic microwave background radiation (CMBR)
implies that the universe since photon decoupling is
almost spatially homogeneous and isotropic.
This is proof of the stability under perturbation of the original
exact
Ehlers--Geren--Sachs (EGS) theorem [2],
and shows that almost--isotropy of the
CMBR is the foundation of the assertion that the universe is
almost a Friedmann--Robertson--Walker (FRW) spacetime. (Indeed
the analysis of the CMBR via the Sachs--Wolfe effect begins with the
{\it assumption} that the universe is almost--FRW since decoupling.)

In this paper we extend the result of [1] to a radiation--dominated
universe where gravity obeys the higher--order
theory with Lagrangian ${\cal L} = R + \alpha R^2$.
In an earlier paper [3] we showed that the exact EGS theorem holds
also in the $\alpha\neq 0$ theory, i.e. if the CMBR is exactly
isotropic for a congruence of freely falling observers, then in
$R + \alpha R^2$ gravity the spacetime is FRW.

The corrections introduced by the $\alpha R^2$ term in the Lagrangian
should be dominant in the very early universe. In fact the
higher--order term can generate inflationary expansion
without invoking an extraneous inflaton field [4,5].
This paper investigates the effect this term may have on massless
particles which decoupled in the early universe. A possible
application is the relic graviton radiation, which decoupled around
the Planck time and has since not interacted significantly
with matter or radiation.
In principle the gravitational background radiation
provides a link to the spacetime geometry of the
very early universe - although in practice any observation of this
background radiation seems a very distant possibility.

This radiation would inevitably contain anisotropies that carry an
imprint
of perturbations at the decoupling time. These anisotropies would be
preserved in the subsequent free propagation of the radiation.
The result proved here
shows that {\em if any gravitational background radiation is observed
in the late universe to be almost isotropic, and if gravity is
governed by the higher--order theory soon after Planck time, then
the spacetime geometry at that time is almost--FRW.} If the
higher--order corrections are neglected, the result holds for
the special case of general relativity ($\alpha=0$), and represents
an extension of [1] from a matter--dominated to a
radiation--dominated universe.

A second application, where the observational possibilities are much
greater,
but the significance of higher--order effects much less,
is a (massless)
neutrino background after decoupling.
Anisotropies in this background would carry information
about the spacetime geometry at $t\approx 1$s.
An estimate of the relative
size of the higher--order terms at neutrino decoupling
depends on the limits placed on the coupling constant
$\alpha$. In [5], a cosmological model is constructed with an
$\alpha R^2$
inflationary epoch, followed by an oscillatory phase (when
re--heating takes place), which leads into the standard Friedmann
radiation
era. In this model, the inflationary constraints (e.g. on density
perturbations) lead to the limits
$$10^{22}~{\rm GeV}^2~<\alpha^{-1}<~10^{26}~{\rm GeV}^2.$$
According to these limits, we calculate that the
relative correction to the Einstein Lagrangian at the start of the
Friedmann era is at most $$\alpha R\approx 10^{-14}.$$
This is effectively negligible, although the
value is model--dependent. It is more realistic to take general
relativity as holding at neutrino decoupling. Then our result for
$\alpha = 0$ shows that an almost isotropic neutrino distribution
implies an almost FRW spacetime at neutrino decoupling.

There may be other massless species that decoupled before neutrinos,
interact very weakly, and have yet to be detected.
Finally, there is also a more mathematical
motivation for proving an
almost--EGS theorem when $\alpha\neq 0$.
We can consider the higher--order Lagrangian as a
`perturbation' of the general relativity Lagrangian, and we show
that the almost--EGS result is `stable'
under this kind of
perturbation, having previously shown [3] that the exact
EGS result is.

For convenience, we summarise the theorems derived in [1--3].
Consider a congruence of freely falling observers in an expanding
universe, measuring freely propagating radiation (massless
particles).\\
{\it (i) EGS Theorem}: If the radiation is exactly isotropic,
then in general relativity the spacetime is exactly FRW [2].\\
{\it (ii) Almost--EGS Theorem}: If the radiation is almost isotropic
in a matter--dominated universe,
then in general relativity the spacetime
is almost FRW [1].\\
{\it (iii) Higher--order EGS Theorem}: If the radiation is exactly
isotropic,
then in $R + \alpha R^2$ gravity the spacetime is exactly FRW [3].\\

In section 4 we prove:\\
{\it (iv) Higher--order Almost--EGS Theorem}: If the radiation is
almost isotropic in a radiation--dominated universe, then in
$R+\alpha R^2$ gravity the spacetime is almost FRW.\\

The proof of (iii) above rests on showing that the scalar curvature
is spatially homogeneous. This follows in [3] from the field equations
only after detailed calculation. Consequently, the proof of (iv)
rests on the requirement that the Ricci scalar curvature is almost
spatially homogeneous. This is the main part
of the generalisation of the general relativity theorem and
is shown in sections 4.1 and 4.2. Consequently we can invoke
theorem (iii) in our proof that the metric is
almost isotropic and spatially homogeneous
in section 4.3. In order to make the discussion as
self--contained as possible, we briefly review in section 2
the necessary facts from the covariant formalism for analysing
small anisotropies in radiation [1], and in section 3 we
summarise the main points from [3] about the higher--order equations.
\\

{\it Notation:} We follow [1]. The metric $g_{ab}$ has signature
$(-,+,+,+)$.
Einstein's gravitational constant and the speed of light in vacuum
are 1. Round brackets on indices denote symmetrisation,
square brackets anti--symmetrisation. $\nabla_a$ is the covariant
derivative
defined by $g_{ab}$. Given a four--velocity $u^a$, the
associated projection
tensor is $h_{ab} = g_{ab} + u_a u_b$, and the
comoving time derivative and spatial gradient are
$$
\dot Q_{a\dots b}\equiv u^c\nabla_cQ_{a\dots b},
$$
$$
^3\nabla_cQ_{a\dots b}\equiv h_c{}^dh_a{}^e\dots h_b{}^f\nabla_d
Q_{e\dots f}
$$
for any tensor $Q_{a\dots b}$.
Given a smallness parameter $\epsilon$, $O[N]$ denotes $O(\epsilon^N)$
and $A\simeq B$ means $A - B = O[2]$.


\section{Covariant analysis of radiation anisotropy}

Following [1], we do
not assume a background metric, but start from the real spacetime
with almost isotropic radiation, and proceed to show that the real
metric is close to an FRW metric. Of course, it follows from the
higher--order EGS theorem (iii) that if the anisotropy vanishes,
then the metric is FRW.

We assume that the universe is radiation--dominated at the time of
decoupling of the massless particles whose distribution is almost
isotropic. The process of decoupling leads to anisotropies in the
decoupled species, and probably also in the other massless or
ultra--relativistic species. After decoupling,
the freely propagating species preserves its anisotropy, while
any anisotropy in the other species is rapidly removed by
collisions.

A unique physical four--velocity $\widehat{u}^a$ is defined as the
unit normal field to the surfaces $\{\bar{\mu}={\rm constant}\}$,
where $\bar{\mu}$ is the energy density of the thermalised
massless (or ultra--relativistic) species. The distribution of the
decoupled species will appear as almost isotropic to `observers'
comoving with $\widehat{u}^a$. This field is irrotational, but in
general not geodesic. Inspection of the proof in [1] and of the
higher--order equations in [3] shows that it is easier to work with
a geodesic but rotating four--velocity $u^a$,
which may be chosen close to $\widehat{u}^a$ on an initial surface.
Then the difference between the
two vectors will remain $O[1]$ provided that the spatial gradient of
the energy density $\bar\mu$ relative to $u^a$ is $O[1]$,
which is proved in section 4 (see (24)).
So the anisotropy measured by freely falling
`observers' comoving with $u^a$
is also small. Thus, although $u^a$ is not uniquely and
physically defined, the $O[1]$ nature of anisotropies relative to
$u^a$, and conclusions based on that, are covariant.

The congruence is necessarily expanding since it must
`track' the expanding $\widehat{u}^a$. Thus $u^a$ satisfies
\be
a^a\equiv\dot{u}^a = 0~,~\Theta\equiv u^a{}_{;a}>0~.
\ee
The rate of expansion
defines a Hubble rate $H$ and average scale factor $S$ by
$\Theta=3H=3\dot{S}/S$. The kinematics of the congruence is determined
by $\Theta$ and the shear $\sigma_{ab}$ and vorticity
$\omega_{ab}$.

Now $u^a$ defines an invariant 3+1 splitting of
tensors
[1,6]. In particular, for a massless particle four-momentum
$$
p^a = E(u^a + e^a)\,, ~~ e_a u^a = 0\,, ~~ e_a e^a = 1\,,
$$
where $E$ is the energy and $e_a$ the direction of
momentum, relative to the `observers' above.
After decoupling, the total radiation distribution function is
$$
f_{tot}(x^c,E,e^d) = \bar f(x^c,E) + f(x^c,E,e^d)
$$
where $\bar f$ is a collision--dominated Planckian
distribution describing the thermalised species, while $f$ is the
distribution function of the decoupled species. This distribution
function may be expanded as [1,7]
\be
f(x^c,E,e^d) = F(x^c,E)+F_a(x^c,E)e^a+F_{ab}(x^c,E)e^a e^b+ \dots
\ee
where the covariant multipole moments $F_{a_1 \dots a_L}(x^c,E)$
for $L \ge 1$ are symmetric trace--free tensors orthogonal to $u^a$,
that
provide a measure of the deviation of $f$ from exact isotropy
(as measured
by $u^a$). If the decoupled radiation is almost isotropic, then [1]:
\be
F,~\dot{F} = O[0] \, ,~~ F_{a_1 \dots a_L},~\nabla_b
F_{a_1 \dots a_L} = O[1] ~~(L \ge 1)\, .
\ee

Energy integrals of the first three moments
define the decoupled
radiation energy density, energy flux and anisotropic
stress:
\bea
\mu = 4\pi \int_0^\infty E^3 F dE = 3p = O[0] \,,\\
q_a = {{4\pi}\over 3} \int_0^\infty E^3 F_a dE = O[1]\,, \\
\pi_{ab} = {{8\pi}\over 15}\int_0^\infty E^3 F_{ab} dE = O[1]~. 
\eea
(In [1] $\mu_R$ is used for $\mu$.) We will also need the integral of
the octopole moment:
\be
\xi_{abc} = {{8\pi}\over 35}\int_0^\infty E^3 F_{abc} dE = O[1]\,.
\ee
The total radiation energy--momentum tensor is
\be
T_{ab} = (\bar\mu + \mu) u_a u_b +{\ts{1\over 3}}
(\bar\mu + \mu)h_{ab}+
\pi_{ab}+2u_{(a}q_{b)}~.
\ee
Since the decoupled and thermalised species are non--interacting,
they separately obey the conservation equations:
\bea
(\bar\mu)^{\displaystyle\cdot}+{\ts{4\over3}}\bar{\mu}\Theta &=&0~,
\\
\dot{\mu} + {\ts{4\over 3}} \mu\Theta +
\pi_{ab}\omega^{ab} +^3\nabla_a q^a &=& 0~, \\
\dot{q}_a +
(\omega_{ab} + \sigma_{ab} +
{\ts{4\over 3}}\Theta h_{ab})q^b+\ts{1\over3}(^3\nabla_a\mu)
+^3\nabla_b\pi^b{}_a &=& 0~, 
\eea
where we have used (1).

After decoupling, the radiation obeys the
equilibrium Boltzmann equation
$$
L(f_{tot}) = 0~,
$$
where $L(\bar f)$ vanishes because of detailed balancing of the
collisions in the thermalised component, and $L(f)$ vanishes because
the decoupled species is collision--free:
\be
L(f) \equiv p^a{\partial f\over\partial x^a}-\Gamma^a{}_{bc}p^bp^c
{\partial f\over\partial p^a}= 0~.
\ee
Consequently $q_a$ and $\pi_{ab}$ in (8) are not dissipative
quantities, but measure the deviation of $f$ from isotropy.
{}From (3), (5) and (6), it follows that
\be
\dot q_a~,~{^3\nabla}_b q_a = O[1]~,~~\dot \pi_{ab}~,~
{^3\nabla}_c \pi_{ab} = O[1]~.
\ee
The Liouville equation (12) may also be covariantly decomposed
into multipole moments.
The monopole moment is (10), the dipole moment is (11),
and the quadropole
moment involves $\xi_{abc}$ and is given in section 4 in
linearised form (see (19)).

\section{Higher--order equations}

The field equations and Ricci and Bianchi identities in higher--order
gravity are given in general 3+1 form in [3].
Quantum corrections to the gravitational Lagrangian which yield the
simplest higher--order field equations are of the form
$$
{\cal L} = R + \alpha R^2~,
$$
where $\alpha$ is a coupling constant. The field equations derived
from this are
$$
R_{ab} -  {\ts{1\over 2}}Rg_{ab} + 2\alpha\left[
R(R_{ab} - {\ts{1\over 4}}Rg_{ab}) + g_{ab}\Box R - R_{;ab}\right] =
T_{ab}~,
$$
where $\Box = g^{ab}\nabla_a\nabla_b$ and $T_{ab}$ is the radiation
energy--momentum tensor (8). Their 3+1 splitting is
\begin{eqnarray*}
R_{ab} u^au^b &=& (1 + 2\alpha R)^{-1}\left[(\bar\mu + \mu) -
{\ts{1\over 2}}R(1 + \alpha R) + 2\alpha \Box R +
2\alpha R_{;ab} u^au^b\right]~, \\
R_{ab} u^ah^b{}_c &=& (1+2\alpha R)^{-1}\left[-q_c +
2\alpha R_{;ab} u^a h^b{}_{c} \right]~,\\
R_{ab} h^a{}_c h^b{}_d &=& (1+2\alpha R)^{-1} \left[
\left\{ {\ts{1\over3}}(\bar\mu + \mu)+{\ts{1\over 2}}R(1 +
\alpha R) -
2\alpha\Box R\right\}h_{cd}+\right. \\
&&\left. +\pi_{cd}+ 2\alpha R_{;ab} h^a{}_c h^b{}_d\right]~,
\end{eqnarray*}
while their trace is (using (8))
\be
R=6\alpha\Box R~.
\ee
Together with the Ricci and Bianchi identities, these give
the constraint and evolution equations governing a radiation spacetime
in higher--order gravity. These equations are given in full in [3],
and will be quoted when needed.

The FRW metric is
$$
ds^2 = -dt^2 + S^2(t)\left[{dr^2\over{(1 - kr^2)}} + r^2d\Omega^2
\right]~,
$$
where $k = 0,\pm 1$. The field
equations for this metric are [5]
\bea
{2\ddot S\over{S}} + {{(\dot S^2 + k)}\over{S^2}} &=&
-{\ts{1\over 3}}(\bar\mu + \mu) -
2\alpha\left[9\left\{{\ddot S\over{S}} +
{(\dot S^2 + k)\over{S^2}}\right\}^2 +
\ddot R + {2\dot S\dot R\over{S}}\right]~, \\
{{3(\dot S^2 + k)}\over{S^2}} &=&  (\bar\mu + \mu) -
2\alpha\left[
9\left\{{(\dot S^2 + k)^2\over{S^4}} - {\ddot S^2\over{S^2}}\right\} +
{3\dot S\dot R\over{S}}\right]~.
\eea
Equation (16) is the higher--order Friedmann equation for radiation
and is
compatible with (15) through the conservation equations (9), (10).
In general relativity,
the FRW spacetime is uniquely characterised by the existence
of a geodesic, expanding four--velocity with zero vorticity and shear.
In higher--order gravity, this is also true for radiation, but
not in general (i.e. not without further conditions) [3].
The four--velocity is then the unique
four--velocity normal to the spatial hypersurfaces and with respect
to which the radiation is exactly isotropic.

\section{Almost--EGS theorem in higher--order gravity}

Following the approach of [1], and given theorem (iii), we must show
that if the deviations from isotropy in $f$ are small, i.e. if (3)
holds, then: (a) all
covariant quantities that vanish in FRW spacetime are small;
(b) the metric can be put into a perturbed FRW form.

It is clear that any covariant quantity which is non--zero in the
exactly isotropic (and therefore FRW) case is an $O[0]$ quantity.
However, if a covariant quantity vanishes in
the exactly isotropic (FRW) case, it is not necessarily $O[1]$. That
is precisely what has to be proven.

We follow the arguments of [1], explaining the deviations in the
proof of the theorem caused by the higher--order equations.
We will only summarise the proof where the equations are independent
of the field equations used, and include details where they are not.\\


\subsection{Almost--FRW kinematics}

The nature of the radiation
kinematic quantities is independent of the
field equations and follows from the covariant multipole
decomposition of the Liouville
equation (12). The zero and first moments are the energy and
momentum conservation equations (10) and (11) respectively.
The momentum
conservation equation (11), together with (13), implies that
\be
{^3\nabla}_a\mu = O[1]~,
\ee
and any of its derivatives are also at least $O[1]$.
The definition of ${^3\nabla}_a$ leads to the following identity [1]:
$$
\left({^3\nabla}_a{^3\nabla}_b - {^3\nabla}_b{^3\nabla}_a\right)
\mu
= -2\omega_{ab}\dot{\mu}~.
$$
By (3) and (17) this implies that
\be
\omega_{ab} = O[1]~.
\ee
The quadrupole moment of (12) is the evolution equation for the
anisotropic
stress tensor $\pi_{ab}$, which has linearised form [1]
\be
\dot{\pi}_{ab}+{\ts{4\over 3}}\Theta \pi_{ab}+
{\ts{8\over 15}}\mu \sigma_{ab} +
2\left\{{^3\nabla}_{<a} q_{b>}\right\}
+ {^3\nabla}_c \xi^c{}_{ab} \simeq 0~,
\ee
where the angled brackets denote the spatially projected, trace--free
and symmetrised part of the enclosed indices, i.e.
$$
S_{<ab>}\equiv \left[ h_{(a}{}^c h_{b)}{}^d -
{\ts{1\over3}}h_{ab}h^{cd}\right]S_{cd}~,
$$
for any $S_{ab}$. Using (3), (13) and (18), (19) implies
\be
\sigma_{ab} = O[1]~,
\ee
and similarly for all of its derivatives. The linearised conservation
equations are thus
\bea
\dot{\mu} + {\ts{4\over 3}}\mu\Theta
+ {^3\nabla}_a q^a &\simeq &0~, \\
\dot{q}_a + {\ts{4\over 3}}\Theta q_a +{\ts{1\over 3}}
\left({^3\nabla}_a \mu\right) +
{^3\nabla}_b \pi^b{}_a &\simeq &0~.
\eea
Taking spatial gradients of (21) and using (4), (5) and (17),
we obtain
\be
{^3\nabla}_a\Theta = O[1]~.
\ee
Equation (23) and the spatial gradient of (9) then lead to
\be
^3\nabla_a \bar{\mu}=O[1]~.
\ee
{}\\
Equation (23) is crucial to proving almost--homogeneity of the scalar
curvature. The higher--order $R_{bc}u^bh^{ca}$ field equation
is the constraint equation [3]
$$
h^{ab}(\ts{2\over 3}\Theta_{;b} - \sigma_{bc;d} h^{cd}) -
\eta^{abcd}u_b\omega_{c;d} =
(1+2\alpha R)^{-1} [q^a - 2\alpha h^{ab}R_{;bc}u^c]~,
$$
where $\omega^a = {\ts{1\over 2}}\eta^{abcd}u_b\omega_{cd}$ and we
have used (1). This equation implies,
by (5), (18), (20) and (23), that
\be
h_a{}^bR_{;bc}u^c = O[1]~.
\ee
Equation (25) is insufficient to show that the spatial gradient of
the scalar curvature is $O[1]$. This requires the
higher--order Raychaudhuri equation [3], whose linearised form is
\be
\dot\Theta + \ts{1\over 3}\Theta^2 +
(1+2\alpha R)^{-1} (\bar\mu + \mu) -
\alpha (1+2\alpha R)^{-1} \left[\ts{1\over 2} R^2 + \Box R -
2 R_{;ab}u^au^b\right]\simeq 0~,
\ee
where we have used (18) and (20).
We take the spatial gradient of (26), and use (14)
together with (17), (18), (20),
(23) and (25), to obtain
\begin{eqnarray*}
\left[\alpha(2\dot{\Theta} + {\ts{2\over 3}}\Theta^2 - R) -
{\ts{1\over 6}}\right]{^3\nabla_c R} + (1 + 2\alpha R)
{^3\nabla_c}(\dot\Theta + {\ts{1\over 3}}\Theta^2) + &&\\
2\alpha\left[h_c{}^dR_{;abd}u^au^b + R_{;ab}
(h_c{}^du^a{}_{;d}u^b + u^ah_c{}^du^b{}_{;d})\right] +
{^3\nabla_c}(\bar\mu+\mu) &\simeq& 0.
\end{eqnarray*}
This implies, using (17), (18), (20), (23), (24) and (25), that
$$
\left[\alpha({\ts{8\over3}}
\dot{\Theta} + {\ts{2\over 3}}\Theta^2 - R) -
{\ts{1\over 6}}\right]{^3\nabla_c R} = O[1].
$$
Since they are non--zero in the exactly isotropic (and thus FRW)
case, $R$, $\Theta$ and $\dot\Theta$ are $O[0]$. Hence we deduce
that
\be
^3\nabla_a R = O[1]~~\mbox{and thus}~~R_{;<cd>} = O[1]~.
\ee
{}\\
The definition of the magnetic part of the Weyl tensor $H_{ab}$ in
terms of the kinematic quantities (18) and (20) is
independent of the field equations so that, as in [1]
\be
H_{ab} = O[1]~.
\ee
However, to prove that the electric part $E_{ab}$ of the Weyl tensor
is $O[1]$ requires the higher--order shear propagation equation [3]:
\bea
&&h_a{}^c h_b{}^d \dot\sigma_{cd}
+\omega_a\omega_b +\sigma_{ac}\sigma^c_{\; b} +\ts{2\over 3}
\Theta\sigma_{ab} -\ts{1\over 3} h_{ab} [\omega^2
+2\sigma^2 ] + \nonumber \\
&&E_{ab} - \ts{1\over 2}(1+2\alpha R)^{-1} \pi_{ab} -
\alpha(1+2\alpha R)^{-1} R_{;<cd>} = 0~,
\eea
on using (1). Then, by (18), (20), (27) and (29)
\be
E_{ab} = O[1]~.
\ee
{}\\
This establishes that the kinematic quantities for the radiation and
geometry are almost--FRW, i.e. {\em all the covariant
quantities that vanish in the exactly isotropic (FRW) case are at
least $O[1]$}. The higher--order terms in the field equations
thus have
no $O[0]$ effect on any of the kinematic quantities.

\subsection{Almost--FRW dynamics}

It follows from section 4.1 that the $O[0]$ equations governing
the dynamics are just those of an FRW universe.
Thus by (26) and the conservation equation (21), the
evolution of the scale factor $S$
is to $O[0]$ that of a higher--order radiation FRW universe.
The first integral of (26) is the almost--homogeneous analogue of
the higher--order Friedmann equation (16)
\be
{3k\over{S^2}} \simeq -3{\dot S^2\over{S^2}} +
(\bar\mu + \mu) - 2\alpha\left[
9\left\{\left({\dot S^2 + k\over{S^2}}\right)^2 -
\left({\ddot S\over{S}}\right)^2\right\} +
{3\dot S\dot R\over{S}}\right],~~~\dot k = O[1]~.
\ee
Then the dynamical equations differ
from those of the exactly isotropic (FRW) metric at $O[1]$ only.
Because the kinematic quantities are just those of a perturbed
FRW universe, it is now possible to linearise the covariant equations
about the FRW values in the usual manner [8],
and to consider the evolution
of density inhomogeneities. The background metric will obey the FRW
equations, and consequently we can use the
usual covariant FRW perturbation analysis.

\subsection{Almost--FRW metric}

By returning to the frame $\widehat{u}^a$ described in section 2,
we now show that the metric
may be given in an almost--FRW form.
In [1], the choice of $\widehat{u}^a$ is based on surfaces
of constant matter density, since for
applications to the CMBR, the universe
is matter--dominated. In our case, $\widehat u^a$ is based on the
surfaces of constant thermalised radiation density
($\widehat{u}_a\propto\bar{\mu},_a$), and
$$
\widehat\omega_{ab} = 0~,~ \widehat a^a\neq 0~.
$$
(All kinematic quantities
of the $\widehat u^a$ congruence are denoted by a hat.)

We are able to choose the geodesic $u^a$ to be close to $\widehat u^a$
because of (24), which ensures that the
angle between $\widehat {u}^a$ and $u^a$ will be $O[1]$. Clearly
the energy flux and anisotropic stress relative to $\widehat{u}^a$
remain $O[1]$.
Changes to all $O[1]$ quantities will be $O[2]$,
but changes
to $O[0]$ quantities will be $O[1]$,
and will introduce effective
energy fluxes with respect to $\widehat {u}^a$.

The congruence $\widehat u^a$ will be normal to cosmic time surfaces
(coinciding with surfaces of constant $\bar\mu$) in
the perturbed model, and the spacetime metric will be [1]:
\be
ds^2 = -A^2(x^a)dt^2 + \widehat h_{\alpha\beta}dx^\alpha dx^\beta~,
\ee
where $\widehat u^a = A^{-1}\delta^a{}_0~,~\widehat h_{ab} =
g_{ab} + \widehat u_a\widehat u_a$, and Greek indices run from 1 to 3.
The kinematic quantities for $\widehat u^a$ are [1]
\be
\widehat a^a\equiv\widehat u^a{}_{;b}\widehat u^b = -\widehat h^{ab}
(\log A),_b~,~~\widehat\omega_{ab} = 0~,~~
\widehat\Theta_{ab} = {\ts{1\over 2}}A\widehat h_{ab,0}~.
\ee
If we express the non--coincidence of the flows of $\widehat u^a$ and
$u^a$ by [1]
\be
\widehat u^a = u^a - V^a + O[1]~~,~~u_aV^a = 0~~,~~V_aV^a = O[1],
\ee
then the kinematic and dynamic quantities
relative to $\widehat u^a$ are given in terms of those relative to
$u^a$ by
\bea
&& \widehat\Theta = \Theta + O[1]~,~~
\widehat\sigma_{ab} = \sigma_{ab} + O[2]~,~~
\widehat a^a = O[1] \nonumber \\
&&\widehat{\mu} = \mu + O[1]~,~~
\widehat{q}^a = q^a + {\ts{4\over 3}}\mu V^a + O[2]~,~~
\widehat{\pi}_{ab} = \pi_{ab} + O[2]~,
\eea
where $\widehat \mu$ is the energy density of the decoupled species
as measured by `observers' comoving with the thermalised species.
Thus the kinematic results (17--30) hold in this frame,
and in particular
\be
\widehat\nabla_a\widehat\mu = O[1]~,~~
\widehat\nabla_a\widehat\Theta = O[1]~.
\ee
Since the expansions are equal (up to $O[1]$), the scale factor
$\widehat S(t, x^\alpha)$, defined by
$$A^{-1}(\log {\widehat S}),_0 =
\ts{1\over 3}{\widehat\Theta}~,$$
will satisfy the higher--order Friedmann equation (31).
{}From (36) and this
definition of $\widehat S$ we can write the spatial metric in (32) as
\be
\widehat h_{\alpha\beta}(t,x^\gamma)dx^\alpha dx^\beta =
\widehat S^2(t,x^\gamma)
\widehat f_{\alpha\beta}(t,x^\gamma)dx^\alpha dx^\beta~,
\ee
where most of the time variation is in $\widehat S$, and
from the result [1]
$$
\widehat\sigma_{ab} = {\ts{1\over 2}}A\widehat S^2f_{ab,0} = O[1]~,
$$
most of the spatial coordinate dependence is in $f_{ab}$.
Equations (33)
and (35) imply that the function $A(t,x^\alpha)$ is only weakly
dependent
on the spatial coordinates, and we can consequently set
$A(t,x^\alpha) = 1 + O[1]$ by rescaling the time coordinate.

It remains only to show that the embedding of the 3--surfaces is
almost--isotropic. The higher--order Gauss--Codazzi equations [3]
for the
surfaces $\{\bar\mu = {\rm const}\}$ give the trace--free part of
the Ricci tensor of these surfaces as
\bea
^3\widehat R_{ab}-\ts{1\over3}(^3\widehat R)
\widehat{h}_{ab} &=&
\widehat h_a{}^c\widehat h_b{}^d\left[\widehat a_{(c;d)} -
\widehat\sigma_{cd;e}\widehat u^e\right] - \widehat\Theta\widehat
\sigma_{ab} +
\widehat a_a\widehat a_b - \ts{1\over 3}\widehat h_{ab}
\widehat{a}^c{}_{;c} + \nonumber \\
&&(1+2\alpha R)^{-1}\left[\widehat\pi_{ab} +
2\alpha(\widehat h_a{}^c\widehat h_b{}^d -
{\ts{1\over 3}}\widehat h_{ab}\widehat h^{cd})R_{;cd}\right]~.
\eea
The scalar curvature is independent of the
four--velocity, and consequently
(27) gives
\be
\left(\widehat h_a{}^c \widehat h_b{}^d -
\ts{1\over 3}\widehat h_{ab} \widehat h^{cd}\right) R_{;cd} = O[1]~.
\ee
Given (39) and
the almost--FRW conditions on the kinematic quantities (35),
equation (38) implies
\be
^3\widehat R_{ab} - {\ts{1\over 3}}(^3\widehat R)
\widehat h_{ab} = O[1]~,
\ee
where, from [3]
\be
^3\widehat R = -{\ts{2\over 3}}\widehat\Theta^2 + (1 + 2\alpha R)^{-1}
\left[ 2\widehat\mu + \alpha R^2 + 4\alpha\widehat h^{ab}R_{;ab}
\right] + O[1].
\ee
Because $R$, $\widehat\Theta$ and $\widehat\mu$ are almost spatially
homogeneous, by (36) and (41), so is $^3\widehat R$. Then the
3--spaces are
spaces of almost--constant curvature:
$$
^3\widehat{R}_{abcd}=\ts{1\over6}(^3\widehat R)
\left[\widehat{h}_{ac}\widehat{h}_{bd}-\widehat{h}_{ad}
\widehat{h}_{bc}\right]+O[1]~.
$$
Thus, as in [1], we have recovered (to
$O[0]$) all of the standard relations governing an FRW universe,
and in
particular, we have shown that there exists an almost--FRW metric in
higher--order gravity, given in almost--comoving coordinates.
\[ \]
We have shown that in a radiation--dominated universe, an
almost--isotropic background radiation implies an almost--FRW
spacetime geometry in the higher--order theory of gravity
(including general relativity as a special case).
In effect, together with the higher--order exact EGS theorem of [3],
we have now shown that the EGS theorem is stable under
perturbations of both the
Lagrangian and the background metric.


\[ \]
{\bf Acknowledgements:}
DRT thanks David Matravers and Portsmouth University for hospitality
and
a research grant. This work was inspired by discussions with Reza
Tavakol.
Thanks to Henk van Elst and Steve Rippl
for helpful discussions. DRT thanks
Queen Mary and Westfield College for
hospitality and support during the completion of this work.
\[ \]
\ni{\bf References}
\[ \]
\ni 1. Stoeger WR, Maartens R and Ellis GFR (1995) {\em Astrophys. J.}
{\bf 443}, 1.\\
\ni 2. Ehlers J, Geren P and Sachs RK (1968) {\em J. Math. Phys.}
{\bf 9}, 1344.\\
\ni 3. Maartens R and Taylor DR (1994) {\em Gen. Rel. Grav.}
{\bf 26}, 599.
({\em Erratum} in (1995) {\em Gen. Rel. Grav.} {\bf 27}, 113.) \\
\ni 4. Barrow JD and Ottewill A (1983) {\em J. Phys. A}
{\bf 16}, 2757.\\
\ni 5. Miji\'c MB, Morris MS and Suen WM (1986) {\em Phys. Rev. D}
{\bf 34}, 2934.\\
\ni 6. Ellis GFR (1971) in {\it General Relativity and Cosmology},
edited by Sachs RK (Academic, New York).\\
\ni 7. Ellis GFR, Matravers DR and Treciokas R (1983) {\em Ann. Phys.}
{\bf 150}, 455.\\
\ni 8. Bruni M, Dunsby PKS and Ellis GFR (1992) {\em Astrophys. J.}
{\bf 395}, 34.\\

\end{document}